\documentclass[letterpaper,journal]{IEEEtran}
\usepackage[T1]{fontenc}
\usepackage{amsmath,amsfonts}
\usepackage{csquotes}
\usepackage[colorlinks=true,urlcolor=blue,citecolor=.,linkcolor=.]{hyperref}
\usepackage[noabbrev]{cleveref}
\usepackage{tabulary}
\usepackage{array}
\usepackage{xspace}
\usepackage{makecell}
\usepackage{url}
\usepackage[export]{adjustbox}
\usepackage[caption=false,font=footnotesize]{subfig}
\usepackage[ruled,linesnumbered]{algorithm2e}
\crefname{algocf}{algorithm}{algorithms}
\Crefname{algocf}{Algorithm}{Algorithms}
\let\oldnl\nl
\newcommand{\nonl}{\renewcommand{\nl}{\let\nl\oldnl}}

\usepackage{comment}
\usepackage[american]{babel}
\usepackage{orcidlink}
\usepackage{tikz}
\newcommand*\circled[1]{\tikz[baseline=(char.base)]{
            \node[shape=circle,draw,inner sep=1pt] (char) {#1};}}
\usepackage{enumitem}
\newlist{rqs}{enumerate}{2}
\setlist[rqs,1]{label=RQ\arabic*.,ref=RQ\arabic*}
\setlist[rqs,2]{label=(\alph*),ref=\thequestionsi(\alph*)}
\def \theosmodelsmall {\texttt{bge-small-en-v1.5}\xspace}
\def \theosmodellarge {\texttt{NV-Embed-v2}\xspace}
\def \theagent {Discovery Agent\xspace}
\def \therag {OpenAPI RAG\xspace}
\newcommand{\inlineparagraph}[1]{\noindent{\textbf{#1.}}}
\usepackage[acronym]{glossaries-extra}
\newabbreviation{rag}{RAG}{Retrieval Augmented Generation}
\newabbreviation{llm}{LLM}{Large Language Model}
\newabbreviation{soc}{SOC}{Service-Oriented Computing}
\newabbreviation{uddi}{UDDI}{Universal Description, Discovery, and Integration}
\newabbreviation{ubr}{UBR}{UDDI Business Registry}
\newabbreviation{nlp}{NLP}{Natural Language Processing}
\newabbreviation{is}{IS}{Information System}
\newabbreviation{ise}{ISE}{Information System Engineering}
\newabbreviation{gics}{GICS}{Global Industry Classification Standard}

\def \thetitle {Retrieval-Augmented Generation for Service Discovery: Chunking Strategies and Benchmarking}
\def \thebenchmark {SOCBench-D\xspace}
\begin{document}

\title{\thetitle}

\author{Robin D. Pesl\orcidlink{0000-0002-5980-9395}, Jerin G. Mathew\orcidlink{0000-0002-4626-826X}, Massimo Mecella\orcidlink{0000-0002-9730-8882}, and Marco Aiello\orcidlink{0000-0002-0764-2124}
\thanks{Robin D. Pesl and Marco Aiello are with the Institute of Architecture of Application Systems, University of Stuttgart, Stuttgart, Germany (email: \{robin.pesl,marco.aiello\}@iaas.uni-stuttgart.de}%
\thanks{Jerin G. Mathew and Massimo Mecella are with the Dipartimento di Ingegneria Informatica, Sapienza Università di Roma, Rome, Italy (email: \{mathew,mecella\}@diag.uniroma1.it)}%
\thanks{Manuscript received March 7th, 2025; revised TBD.}}

\markboth{}%
{R.D. Pesl \MakeLowercase{\textit{et al.}}: \thetitle}

\maketitle
\begin{abstract}
Integrating multiple (sub-)systems is essential to create advanced Information Systems.
Difficulties mainly arise when integrating dynamic environments, e.g., the integration at design time of not yet existing services.
This has been traditionally addressed using a registry that provides the API documentation of the endpoints.
\glsxtrfullpl{llm} have shown to be capable of automatically creating system integrations (e.g., as service composition) based on this documentation but require concise input due to input token limitations, especially regarding comprehensive API descriptions.
Currently, it is unknown how best to preprocess these API descriptions.
In the present work, we (i)~analyze the usage of \glsxtrfull{rag} for endpoint discovery and the chunking, i.e., preprocessing, of state-of-practice OpenAPIs to reduce the input token length while preserving the most relevant information.
To further reduce the input token length for the composition prompt and improve endpoint retrieval, we propose (ii)~a \theagent that only receives a summary of the most relevant endpoints and retrieves specification details on demand.
We evaluate \glsxtrshort{rag} for endpoint discovery using (iii)~a proposed novel service discovery benchmark \thebenchmark representing a general setting across numerous domains and the real-world RestBench benchmark, first, for the different chunking possibilities and parameters measuring the endpoint retrieval accuracy.
Then, we assess the \theagent using the same test data set.
The prototype shows how to successfully employ \glsxtrshort{rag} for endpoint discovery to reduce the token count.
Our experiments show that endpoint-based approaches outperform na\"ive chunking methods for preprocessing.
Relying on an agent significantly improves precision while being prone to decrease recall, disclosing the need for further reasoning capabilities.
\end{abstract}


\section{Introduction}
\label{sec:intro}

\IEEEPARstart{T}{he} specification language {OpenAPI} is the state-of-practice for describing interfaces for integrating systems.
It contains formal elements like paths and natural language constituents such as descriptions.
For integrating these systems automatically, automated service composition using \glspl{llm} has been recently proposed~\cite{pesl2024verfahren,pesl2024uncovering,pesl2024compositio}.
These approaches exploit the capabilities of \glspl{llm} to process formal and natural language input, combining them with the inherent nature of automated service composition of decoupling and independent lifecycle management.
While saving on manual modeling efforts by relying on already broadly available OpenAPIs, the approaches face the challenge of limited input token length~\cite{pesl2024compositio}.
This bounds the quantity and extent of the input service description.
Even for proprietary models with a large input token context, e.g., OpenAI's GPT4 with a context size of 128,000 tokens~\cite{openai2024contextsize}, an economic constraint emerges as these use of the models is paid in relation to the input and output token count.
Therefore, a smaller prompt length is beneficial both for inserting further service documentation and reducing proprietary models' usage costs.

To address these challenges, \gls{rag}~\cite{lewis2020retrieval} has emerged as a promising technique.
In such an approach, the external information is collected in a database, typically structured as a set of documents or document chunks.
The primary goal is retrieving only a small subset of the most relevant documents or document chunks, which is then inserted into the prompt~\cite{lewis2020retrieval}.
How to optimally apply \gls{rag} for endpoint discovery is open to investigation, leading to the following research questions that we address in this paper:
\begin{rqs}[topsep=0pt,align=left,wide=0pt,leftmargin=*]
    \item How to benchmark service discovery with natural language queries across the most relevant domains?
    \item How best to preprocess, i.e., chunk, OpenAPIs for \gls{rag} endpoint discovery?
    \item Can \gls{llm} agents be employed to reduce token count further and improve retrieval performance?
\end{rqs}

For answering RQ1 and extending our previous work~\cite{pesl2025socrag}, we propose the novel service discovery benchmark \thebenchmark comprising natural language query $q$, expected endpoints $e_\text{expected}$ pairs to evaluate \gls{rag} for OpenAPI endpoint discovery thoroughly.
We rely on the \gls{gics}~\cite{gics2023} as a leading standard for classifying industries into sectors, i.e., domains, to ensure generalizability across various domains.
It provides the following domains: energy, materials, industrials,  consumer discretionary,  consumer staples,  health care, financials, information technology,  communication services, utilities, and real estate.
Similar to the ToolBench approach~\cite{qin2023toolllm}, which employs ChatGPT to create training data, we use an \gls{llm} to construct for each of the domains five services with ten endpoints each as OpenAPIs.
We validate the OpenAPIs syntactically using an OpenAPI validator and semantically using another \gls{llm}.
Using the services for each domain, we let an \gls{llm} create ten queries for a random subset $e_\text{expected}$ of the endpoints, i.e., $e_\text{expected}$ is the solution to $q$.
We recheck the solution correctness using an \gls{llm} and preclude ambiguity between the queries related to the same domain by defining a similarity threshold relying on an embedding model.
To reduce the influence of randomness, we create five instances of the benchmark, resulting in $5 \text{ (benchmark instances)} \cdot 11 \text{ (domains)} \cdot 10 \text{ (queries)} =550$ queries in total.
Based on \thebenchmark, we compute the accuracy metrics across all domains to determine the accuracy and generalizability of the approach.

To answer RQ2, we develop an \textit{\therag} system that takes as input service descriptions.
We apply various token-based and \gls{llm}-based chunking strategies to split the documentation and evaluate them based on retrieval quality.
The token-based strategies process the document using a classical parser and then split the parts into equal-sized chunks.
The \gls{llm}-based strategies let an \gls{llm} create a description, i.e., a summary or a question, for each endpoint and then use these descriptions for similarity matching.
We employ mainstream open-source and proprietary embedding models for similarity matching, which can create an embedding vector for an input.
The similarity between two inputs can then be determined by comparing their embedding vectors using, e.g., the cosine similarity.
We evaluate the \therag and the chunking strategies by relying on our novel \thebenchmark benchmark and the already available RestBench benchmark for \glspl{llm} agents~\cite{song2023restgpt}, measuring recall and precision for each chunking strategy.
We employ \thebenchmark to retrieve generalizable results across multiple domains and the RestBench benchmark consisting of the Spotify and TMDB OpenAPI descriptions and corresponding queries, each with a set of endpoints as the sample solution, for real-world applicability.

To address RQ3, we propose an \gls{llm} agent called \textit{\theagent}.
As \gls{llm} agents allow the usage of external tools, we investigate using one tool that filters and enters the \gls{llm} endpoint summaries to the prompt using \gls{rag}, while the second tool allows the retrieval of the endpoint details on demand.
We resort to the same benchmarks for evaluation and measure recall and precision.
As the chunking strategy, we rely on the \gls{llm}-based summary strategy with OpenAI's \texttt{text-embedding-3-large} embedding model~\cite{openai2024embeddings}.

The remainder of the paper is structured as follows.
First, we provide an overview of related works regarding service discovery and \glspl{llm} in \Cref{sec:related_work}.
Then, we present how to use \gls{rag} for endpoint discovery and the OpenAPI chunking strategies in \Cref{sec:design}.
We introduce \thebenchmark and evaluate and discuss the \therag and the different chunking strategies in \Cref{sec:evaluation}.
Final considerations are presented in \Cref{sec:conclusion}.

\section{State of the Art}
\label{sec:related_work}

Service discovery has been actively investigated in the fields of networking and information systems.
Next, we provide a brief review of the state of the art in that field and an exploration of recent trends in service discovery, including \glspl{llm} and \gls{llm} agents.

\subsection{Service Discovery}

The most common service discovery implementation is based on service registries, which collect information about available services and offer search facilities.
The service registry is usually backed by a component residing at the middleware or application levels~\cite{lemos2015web}.
It is characterized by the syntax used to describe the services and their invocation and the expressive power of the available query language.
The typical integration model is a pull model where service consumers search for the required services.
Some standards, such as UPnP, are based on a push model, where service providers regularly advertise their services~\cite{santana2006upnp}.

In the early days of XML-based Web services, the infrastructure for service discovery was the \gls{uddi} specification~\cite{curbera2002unraveling}.
\gls{uddi} had a global incarnation called the \gls{ubr}, intended to offer an Internet-wide repository of available web services and promoted by IBM, Microsoft, and SAP.
Unfortunately, \gls{ubr} never gained widespread adoption and was short-lived (2000-2006).
Significant research in the early days focused on enhancing service discovery on \gls{uddi}, improving search capabilities, and creating federated registries, e.g.,~\cite{baresi2006distributed,bohn2008dynamic,10.1007/978-3-540-24593-3_5}.
Alternatively, WS-Discovery is a multicast protocol that finds web services on a local network.

Nowadays, OpenAPI is the de facto standard for describing services.
While not offering a discovery protocol and mechanism, given its popularity, OpenAPI would also benefit from discovery~\cite{10.1007/978-3-031-57853-3_3}.
Several authors have proposed the use of additional infrastructure for discovery in the form of centralized repositories (SwaggerHub or Apiary), service registry integration (Consul, Eureka), API Gateways (Kong, Apigee), or Kubernetes annotations (Ambassador).

Populating registries of services requires effort from service providers, which often hinders the success of such approaches, especially if the service provider is expected to provide extensive additional information beyond the service endpoints.
This additional effort has often been the reason for the failure of some of these technologies, most notably \gls{ubr}.
Approaches confined to specific applications, domains, or enterprises have been more successful, e.g., Eureka.
Developed by Netflix as part of its microservices architecture~\cite{thones2015microservices}, Eureka helps clients find service instances described by host IP, port, health indicator URL, and home page.
Developers can add optional data to the registry for additional use cases.

While classical incarnations like \gls{uddi} used to be comprehensive, they required extensive modeling, e.g., as semantic annotations.
To avoid falling into the same pit, our approach proposed here relies on already broadly available OpenAPI specifications.

\subsection{\glsfmtlongpl{llm}}
\glspl{llm} represent one of the recent advancement in the \gls{nlp} and machine learning field~\cite{achiam2023gpt,llama3modelcard,kim2024leveraging}.
Often containing billions of parameters, these models are trained on extensive text corpora to generate and manipulate text with human-level proficiency~\cite{radford2019better}.
They are primarily based on an encoder-decoder architecture called Transformers~\cite{vaswani2017attention}, which has been further refined to improve text generation tasks using decoder-only models such as GPT~\cite{radford2018improving}.
Usually, the input is a natural language task called prompt, which first needs to be translated into a sequence of input tokens.
The model processes the prompt and returns an output token sequence, which can then be translated back to a natural language answer.
As these models have shown the ability to capture intricate linguistic nuances and even semantic contexts, they can be applied to a wide range of tasks, including in software engineering~\cite{fan2023large}.
\glspl{llm} can be used to create integration based on endpoint documentation automatically~\cite{pesl2024verfahren,pesl2024uncovering,pesl2024compositio}.
Yet, these face strict input token limitations, e.g., 128,000 tokens for current OpenAI models~\cite{openai2024contextsize,pesl2024compositio}.

Another approach is encoder-only models such as BERT~\cite{devlin2019bert}, often referred to as embedding models.
They allow condensing the contextual meaning of a text into a dense vector, termed embedding.
Using similarity metrics such as dot product, cosine similarity, or Euclidean distance allows for assessing the similarity of two input texts.
Embedding models are usually used for the similarity search in \gls{rag} systems~\cite{cuconasu2024power}, a technique we also exploit in our implementation.

In previous work, we proposed Compositio Prompto as an architecture to employ \glspl{llm} to automatically generate a service composition as code based on service documentation, a natural language task, and an input and output schema.
A concern was due to the limited input token count of the \gls{llm} and that the \gls{llm} generated imperfect results requiring further manual effort to make the code operational~\cite{pesl2024compositio}.
In the present work, we analyze the usage of \gls{rag} to alleviate the limited input token count issue.

\subsection{\glsfmtshort{llm} Agents}

\glspl{llm} have shown remarkable capabilities in solving complex tasks by decomposing them in a step-by-step fashion~\cite{wei2022chain} or by exploring multiple solution paths simultaneously~\cite{yao2024tree}.
Typically, these plans are generated iteratively by using the history of the previously generated steps to guide the generation of the next step.
Recent studies have shown the potential of providing \glspl{llm} access to external tools to boost their inference capabilities and add further knowledge.
Such an approach consists of prompting the \gls{llm} to interact with external tools to solve tasks, thus offloading computations from the \gls{llm} to specialized functions.
Notable examples of such tools include web browsers~\cite{nakano2021webgpt}, calculators~\cite{cobbe2021training}, and Python interpreters~\cite{gao2023pal}.
In practice, this can be realized as a Python function called during the interaction with the \gls{llm}.

The \gls{llm} agent paradigm~\cite{mialon2023augmented,openai2024function,yao2023react} combines the concepts of external tool usage,  the planning capabilities of \glspl{llm}, and adds a shared memory to solve complex tasks.
Given an input task, an \gls{llm} agent uses its reasoning capabilities to decompose the task into a set of simpler subtasks.
For each subtask, the \gls{llm} finds and interacts with the set of tools to solve the subtask.
Then, based on the outcome of the current task and the history of previously executed subtasks, the \gls{llm} agent generates a new subtask and repeats the steps mentioned above or terminates if the original task is solved.
To instruct the processing, the outcome of the tool invocations and the history of the subtasks are stored in the memory, typically consisting in the \gls{llm} agent's own context.
Within this work, we apply the \gls{llm} agent paradigm to create the \theagent as an \gls{llm} agent for endpoint discovery.

A critical challenge for \gls{llm} agents is the accessibility to a set of common APIs and tasks for their evaluation, e.g., tested using benchmarks like API Bank~\cite{li2023apibank} or RestBench~\cite{song2023restgpt}.
API Bank is a benchmark consisting of a set of APIs exposed through a search engine.
Unfortunately, the available code of the benchmark is incomplete, i.e., all APIs, but only a few of the used queries are available.
The RestBench benchmark contains a collection of tasks and endpoints expressed using the OpenAPI specification of Spotify and TMDB~\cite{song2023restgpt}.
We employ RestBench to validate our results, given that it is the most extensive benchmark available.

OpenAPIs within \gls{llm} agents have been used in RestGPT~\cite{song2023restgpt} and Chain of Tools~\cite{shi2024chain}.
The former combines multiple \gls{llm} agents to solve complex tasks by interacting with a set of tools exposed using the OpenAPI specification. The latter solves an input query by framing the problem as a code generation task and interacts with the set of tools to generate Python code.
In contrast, our \theagent does not directly interact with the endpoints found in the OpenAPIs.
Instead, it filters and returns matching endpoints that can be used for subsequent processing.

Even when considering the similarity to the tool selection within \gls{llm} agents, the task of selecting a set of tools from a larger pool to solve a specific problem remains relatively underexplored~\cite{yuan2024craft}.
Existing research primarily focuses on the a priori selection of human-curated tools~\cite{parisi2022talm}, heuristic-based methods for tool selection~\cite{liang2024taskmatrix}, choosing the relevant tool by scoring each query against every tool using a similarity metric between user queries and API names~\cite{patil2023gorilla}, and embedding-based semantic retrieval using a combination of different vector databases~\cite{yuan2024craft}.
With our work, we contribute the analysis of preprocessing OpenAPIs into this corpus.

\section{\therag}
\label{sec:design}

We first introduce the general architecture to employ \gls{rag} for endpoint discovery.
Then, we investigate how to chunk OpenAPIs as preprocessing for \gls{rag}.

\subsection{\glsfmtshort{rag} for Endpoint Discovery}

\gls{rag} comprises a preprocessing step ahead of the answer generation of an \gls{llm} to enrich the prompt with additional data.
Therefore, a retrieval component performs a semantic search based on some knowledge sources.
Usually, the semantic search is done by embedding similarity, and the data from the knowledge sources is reduced to small chunks to allow fine-grained information retrieval~\cite{lewis2020retrieval}.

\begin{figure*}
    \centering
    \includegraphics[width=0.9\linewidth]{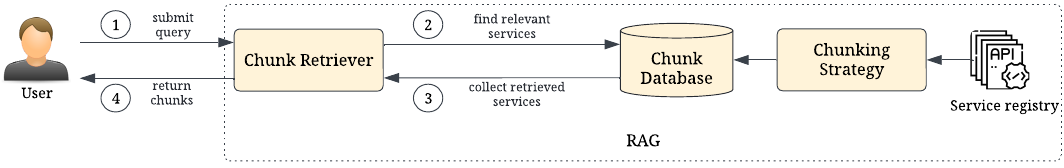}
    \caption{\glsxtrshort{rag} for Endpoint Discovery}
    \label{fig:soc-rag}
\end{figure*}

The application of \gls{rag} for endpoint discovery, i.e., the \therag, is shown in \Cref{fig:soc-rag}.
Initially, the chunking strategy determines how the chunks are created from the OpenAPIs, i.e., how many chunks are created and what they contain.
Each chunk has an embedding as metadata for similarity search in addition to its content.
The chunking strategy specifies which data is used as input to the embedding model to create the embedding.
This input does not have to match the chunk content, e.g., it can be a summary instead of the entire content.
The chunks are finally stored in the chunk database.

For retrieval, the user submits in~\circled{1} a natural language query $q$ to the chunk retriever, which converts $q$ into the embedding $e$ using the same embedding model as for the chunk creation.
In~\circled{2}, the chunk retriever queries the chunk database using $e$.
The chunk database compares $e$ using a similarity metric with the embeddings of the service chunks contained in the database.
The results are the top $k$ most similar chunks according to the metric, which are then returned to the chunk retriever in~\circled{3}.
Finally, in~\circled{4}, the chunk retriever forwards the retrieved results to the user, who can add them to their prompt either manually or automatically through integration into their tooling.

The benefit of employing \gls{rag} is the insertion of only the gist of the available information, which allows picking only the most relevant information for the fix \gls{llm} context size.
A drawback is that, based on the retrieval algorithm, not all relevant information may be retrieved.
Further, fixing $k$ reveals the advantage of controlling the result size.
An alternative is to return all chunks about a certain similarity threshold, introducing the question about the optimal cutoff.

\begin{figure*}
    \centering
    \includegraphics[width=0.9\linewidth]{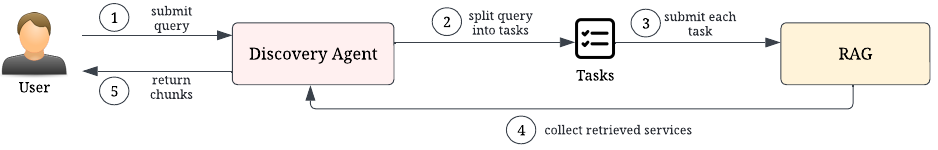}
    \caption{Overview of the \theagent Approach for Endpoint Discovery}
    \label{fig:soc-agent}
\end{figure*}

\Cref{fig:soc-agent} shows how the \theagent extends on the \gls{rag} from \Cref{fig:soc-rag} shown in yellow hued.
Instead of passing $q$ to the \gls{rag}, the user submits it in~\circled{1} to the \theagent, which then iteratively decomposes $q$ into a set of fine-grained tasks in~\circled{2}.
Breaking down the query into smaller, more manageable tasks can potentially fill the gap between the coarse semantics of the query and the specificities in the services documentation.
In~\circled{3}, the \theagent submits each task to the \gls{rag} to retrieve the set of relevant chunks to solve the current task specifically.
Finally, in~\circled{4}, the \theagent collects the retrieval results of each individual task, filters them, and repeats~\circled{2} if $q$ needs further processing or returns the results to the user in~\circled{5}.

\subsection{OpenAPI Chunking Strategies} \label{sec:chunking_strategies}

A critical step in the \gls{rag} workflow is creating the chunks for the chunk database.
Embedding models typically have a limited input token size, and real-world service registries can contain tens of thousands of services, each containing multiple potentially lengthy endpoints due to detailed descriptions or extensive input and output schemas.
A single service might not fit into the context size of the embedding model or even exceed the limit of the \gls{llm} that further processes the output of the \gls{rag} system.
In addition, service documentation can also feature additional metadata that, while valuable for understanding service details, is not necessarily relevant for composing services.

\begin{table}
    \caption{Implemented Chunking Strategies}
    \label{tab:approaches}
    \begin{tabulary}{\columnwidth}{l|l|l|L}
        \textbf{Category} & \textbf{Splitting} & \textbf{Refinement} & \textbf{Meta-Parameters} \\
        \hline
        Token-based & No split & Token chunking & $m$~(model), $s$~(chunk size), $l$~(overlap) \\
                    & JSON split & Token chunking & $m$~(model), $s$~(chunk size), $l$~(overlap) \\
                    & Endpoint split & Token chunking & $m$~(model), $s$~(chunk size), $l$~(overlap) \\
                    & Endpoint split & Remove examples & $m$~(model) \\
                    & Endpoint split & Relevant fields & $m$~(model) \\
                    & Endpoint split & \makecell[tl]{JSON split\\token chunking} & $m$~(model), $s$~(chunk size), $l$~(overlap) \\
        \hline
        \glsxtrshort{llm}-based & Endpoint split & Query & $m$~(model) \\
                                & Endpoint split & Summary & $m$~(model) \\
                                & Endpoint split & CRAFT & $m$~(model)
    \end{tabulary}
    \centering
\end{table}

To determine advantageous chunking strategies, we employ the nine well-known chunking strategies presented in \Cref{tab:approaches}.
Input is always an OpenAPI specification, and output is a list of chunks.
The chunking strategies can be categorized into \textit{token-based} and \textit{\gls{llm}-based} strategies.
Each strategy consists of a \textit{splitting} method, which dissects the OpenAPI specification into a list of intermediate chunks, and another \textit{refinement} step converts the intermediate chunks to the final list of chunks.
In addition, there is the meta-parameter for the used embedding model~$m$.
For the refinement step, there is also the chunk size~$s$ in tokens and their overlap~$l$, i.e., how many tokens two consecutive chunks share.

For the token-based approaches, we consider three main splitting methods.
The \textit{no split} method returns a single intermediate chunk for each OpenAPI containing the whole specification.
The \textit{endpoint split} divides the OpenAPI into one chunk per endpoint.
The \textit{JSON split} is a built-in LlamaIndex\footnote{\url{https://github.com/run-llama/llama_index}\label{foot:llamaindex}} splitting strategy tailored to JSON files.
This strategy parses the JSON file and traverses it using depth-first search, collecting leaf nodes, i.e., key-value pair where the value is a primitive type, e.g., strings, numbers, etc.
During this traversal, the parser concatenates keys and values into single lines of text to create a comprehensive representation of each leaf node.

For the refinement, we implemented \textit{token chunking}, \textit{remove example}, \textit{relevant field}, and \textit{JSON split token chunking}.
The \textit{token chunking} splits each intermediate chunk into a list of fixed-size chunks of $s$ tokens respecting an overlap of $l$ tokens with the previous node.
The \textit{remove example} removes the \texttt{requestBody} and recursively all \texttt{examples} fields for each endpoint as these are typically lengthy but contribute little information.
The \textit{relevant field} extracts representative fields, i.e.,  title, service description, endpoint verb, endpoint path, and endpoint description, which contribute information but few tokens.
In the \textit{JSON split token chunking}, we integrate the JSON split for a single endpoint with subsequent chunking.

For the \gls{llm}-based processing strategies, we apply the endpoint split and a \textit{summary} (similar to~\cite{nogueira2019document}) and \textit{query} approach for refinement.
In the \textit{summary} approach, we prompt an \gls{llm} to generate a summary for each OpenAPI endpoint.
For the \textit{query} approach, we instruct the \gls{llm} to generate a possible query matching the OpenAPI endpoint, as this might be closer to a possible input query than the summary.
As an advanced \gls{llm}-based approach, we implement CRAFT~\cite{yuan2024craft}, which combines multiple retrieval strategies.
For all three approaches, we only consider the \gls{llm} output for the embedding creation.
The chunk content remains the original OpenAPI endpoint information.
The no split and JSON split methods can only be used with token chunking since all other refinement strategies rely on exactly one endpoint as an intermediate chunk.

\section{Evaluation}
\label{sec:evaluation}

To evaluate the \therag and the \theagent, we first implement it as a fully operational prototype.
Then, we create the \thebenchmark to evaluate \gls{rag} across all domains.
Additionally, we employ the RestBench~\cite{song2023restgpt} benchmark to validate the prototype in a real-world setting.

\subsection{Implementation}
We implement the \therag and \theagent approaches as open-source prototypes\footnote{Code: \url{https://github.com/IAAS/SOCBench}, Data: \url{https://dx.doi.org/10.21227/vdm4-k186}\label{foot:sources}} based on the LlamaIndex library.\textsuperscript{\ref{foot:llamaindex}}
For the prototypes, we rely solely on OpenAPIs as the state-of-practice for service descriptions.

For the \therag, we focus on the components presented in \Cref{fig:soc-rag}.
When starting, the system loads the OpenAPIs and applies a chunking strategy to create chunks and their embeddings for their later retrieval.
The chunks contain thereby the information from the OpenAPIs, e.g., a whole endpoint or a part of it.
A chunk embedding does not necessarily have to match the chunk's content; for example, the content can be the endpoint, and the embedding is created using a natural language summary of the endpoint.
Thus, the matching is performed based on the embedding, and the result returned is the chunk's content, which can include additional information not required for the matching process.
As the service database, we use FAISS, which allows the storage and the similarity search of chunks~\cite{douze2024faiss}.
We use a so-called QueryEngine from LlamaIndex for the chunk retriever, which allows us to query a chunk database based on textual input.

As a more advanced algorithm, we implement the retrieval approach from CRAFT~\cite{yuan2024craft}, which utilizes a multi-view matching approach to match tasks to tools.
Transferred to our use case, we employ the result sets of the summary, the endpoint name, and the endpoint description approaches.
If an endpoint appears in at least two of these result sets, it is included in the final result.
We adapt CRAFT to exactly return $k$ results by iterating over the result sets and adding one element after the other.
Therefore, we create an intermediate set for each of the three approaches and successively add the result elements, i.e., first, add the first element of the summary results, then the first element of the endpoint name results, and so on.
After adding an element, we check how many elements are in at least two of the intermediate sets and continue adding elements until there are $k$ elements.

To enable measuring the retrieved endpoints, we attach the endpoint information, i.e., verb and path, to each chunk as metadata.
For the endpoint split splitting strategies, we take the information from the endpoint.
For the other strategies, we first attach a list of all endpoints to the nodes before splitting and then filter on the endpoint paths in the final chunks after splitting.
So, for each chunk, we know to which endpoint or endpoints it relates to.

We realize the \theagent from \Cref{fig:soc-agent} using a LlamaIndex OpenAIAgent, which implements the \gls{llm} agent pattern for OpenAI's \glspl{llm}.
An OpenAIAgent takes a list of tools, i.e., Python functions with a name and a description as parameters, and interacts with these using the OpenAI API.
For the tools, we use a \gls{rag} with chunks of the endpoint's verb, path, and summary as contents and for their embeddings.
We create the summary by instructing an \gls{llm} to create it based on the endpoint information, i.e., as in the summary chunking strategy.
This should reduce the token count, as the chunks are much smaller because not all endpoint details are returned and processed.
To provide all information, we introduce a second tool, which takes the endpoint verb and path as input parameters and returns the whole endpoint information.
The complete data is only inserted into the history for indispensable endpoints.

\subsection{\thebenchmark} \label{sec:socbenchd}
To evaluate the \gls{rag} implementation for service discovery in a generalized setting across various domains, we propose and implement our benchmark \textit{\thebenchmark} based on the \gls{gics}, which comprises all relevant industry domains grouped into eleven sectors.
For each domain of the eleven \gls{gics}, we employ an \gls{llm} to create five services as OpenAPIs, each with ten endpoints.
Using the services within the same domain, we then select ten random subsets of the endpoints and let the \gls{llm} create a natural query for each.
To ensure quality control, we ensure syntactical validity via schema compliance and check semantics by employing another \gls{llm}.
To reduce the influence of randomness, we generate five benchmark instances leading to 50 queries per domain and 550 queries in total.

As the \gls{gics} is designed to encompass all industry sectors, we can also assume to cover all industry domains and therefore generalizability with \thebenchmark.
Further domains are just subdomains of the \gls{gics} sectors.
Therefore, by employing \thebenchmark, we can gain insights on service discovery across domains.

\inlineparagraph{Implementation Details}
\Cref{alg:create_benchmark,alg:create_queries} describe the benchmark creation in detail as pseudo code.
\Cref{alg:create_benchmark} comprises the benchmark creation (\texttt{createBenchmark}), the generation of services (\texttt{createServices}), the endpoint creation (\texttt{createEndpoints}), and the OpenAPI generation (\texttt{createOpenAPI}).
\Cref{alg:create_queries} presents the query creation (\texttt{createQueries}, \texttt{createQuery}) and the semantic endpoint checking (\texttt{checkNecessary}).

First, we call \texttt{createBenchmark}~(1) with the list of domains (\textit{domains}), the number of services $n_s=5$, the number of endpoints each service should contain $n_e=10$, and the number of queries that should be created per domain $n_q=10$ as parameters to create a single benchmark instance.
Hence, to create the five benchmark instances, we invoke \texttt{createBenchmark} five times.
We define the list \textit{benchmark} to collect the services and the pairs of natural language query and expected endpoints, i.e., the OpenAPIs and the queries~(2).
For each of the domains~(3), we collect the OpenAPIs in \text{openapis}~(4), all endpoints of all services in $e_\text{all}$~(5), and create the list of $n_s$ services, i.e., the service name and its description, stored in \textit{services} by calling \texttt{createServices}~(6).
For each of the \textit{services}, we create $n_e$ endpoints, i.e., a list of verb endpoint description triplets, by calling \texttt{createEndpoints}~(8) and add the \textit{endpoints} to all endpoints $e_\text{all}$~(9).
Based on the list of endpoints \textit{endpoints}, we create the service's OpenAPI by invoking \texttt{createOpenAPI}~(10) and add the generated OpenAPI to the list of all OpenAPIs (\textit{openapis}) in the domain~(11).
Given the complete list of OpenAPIs in the current domain, we create $n_q$ \textit{queries}, i.e., the natural language queries and their list of expected endpoints~(13).
We finalize the current domain by adding the \textit{openapis} and \textit{queries} to the \textit{benchmark}~(14).
Once done with all domains, we return the \textit{benchmark} list as the current benchmark instance.
In case of a (validation) error, the current results are stored as files, i.e., we can continue our algorithm from the last valid state.

For \texttt{createServices}~(19), we query the \gls{llm} to return the list of service names and descriptions~(20).
Then, we assert that the correct number of services was returned~(21).
Otherwise, we recreate the list.
Finally, we return the created list~(22).
Equivalent to \texttt{createServices}, in \texttt{createEndpoints}~(25-29), we create the list of endpoints, i.e., the verb, endpoint, and description triplets, which the service should contain.

For the OpenAPI generation based on the list of \textit{endpoints} in \texttt{createOpenApi}~(32), we first query the \gls{llm} to create the OpenAPI~(33).
Then, we validate that exactly the endpoints from \textit{endpoints} are contained in the OpenAPI~(34), followed by a formal verfication\footnote{\url{https://github.com/python-openapi/openapi-spec-validator}} of the OpenAPI ensuring syntactical validity~(35).
Finally, we analyze semantics by prompting the \gls{llm} to evaluate whether the OpenAPI is valid, reasonable, and specific for the domain~(36).
In case of any (validation) errors, we prompt the \gls{llm} with the OpenAPI and the error message to fix the error.
If the \gls{llm} cannot fix the error, we discard the OpenAPI and restart from~(33).

\smallskip
\noindent
\begin{minipage}{\linewidth}
\makeatletter
\@twocolumnfalse
\makeatother
\begin{algorithm}[H]
\caption{Create Benchmark}\label{alg:create_benchmark}
\DontPrintSemicolon
\SetKw{KwTo}{in}
\SetKw{KwAssert}{assert}
\SetKwComment{Comment}{$\triangleright$\ }{}
\SetCommentSty{textrf}
\SetFuncSty{textrf}
\SetNlSty{}{\color{gray}}{}
\SetKwFunction{createBenchmark}{createBenchmark}
\SetKwFunction{createServices}{createServices}
\SetKwFunction{createEndpoints}{createEndpoints}
\SetKwFunction{createOpenApi}{createOpenApi}
\SetKwFunction{createQueries}{createQueries}
\SetKwFunction{queryLLM}{queryLLM}
\SetKwFunction{len}{len}
\SetKwFunction{openApiValidator}{openApiValidator}
\SetDataSty{itshape}
\SetKwData{Benchmark}{benchmark}
\SetKwData{Openapi}{openapi}
\SetKwData{Openapis}{openapis}
\SetKwData{Services}{services}
\SetKwData{Endpoints}{endpoints}
\SetKwData{Queries}{queries}
\SetKwProg{Fn}{function}{}{end}
\Fn{\createBenchmark{domains$,n_s,n_e,n_q$}}{
    \Benchmark $\gets$ []\;
    \For{domain \KwTo domains}{
        \Openapis $\gets$ []\;
        $e_\text{all}\gets$ [] \Comment*[r]{all endpoints}
        \Services $\gets$ \createServices{domain, $n_s$}\;
        \For{service \KwTo services}{
            \Endpoints $\gets$ \createEndpoints{$n_e$}\;
            $e_\text{all}$.extend(\Endpoints)\;
            \Openapi $\gets$ \createOpenApi{\Endpoints}\;
            \Openapis.append(\Openapi)\;
        }
        \Queries $\gets$ \createQueries{\Openapis, $e_\text{all}$, $n_q$}\;
        \Benchmark.append((\Openapis, \Queries))
    }
    \KwRet \Benchmark
}
$t_s\gets$ template create services\;
\Fn{\createServices{domain, $n_s$}}{
    \Services $\gets$ \queryLLM{$t_s$, domain, $n_s$}\;
    \KwAssert \len(\Services) = $n_s$\;
    \KwRet \Services\;
}
$t_e\gets$ template create endpoints\;
\Fn{\createEndpoints{domain, $n_e$}}{
    \Endpoints $\gets$ \queryLLM{$t_e$, domain, $n_e$}\;
    \KwAssert \len(\Endpoints) = $n_e$\;
    \KwRet \Endpoints\;
}
$t_o\gets$ template create openapi\;
$t_c\gets$ template check openapi\;
\Fn{\createOpenApi{endpoints}}{
    \Openapi $\gets$ \queryLLM{$t_o$, \Endpoints}\;
    \KwAssert \Endpoints $=$ \Openapi.endpoints\;
    \KwAssert \openApiValidator{openapi}\;
    \KwAssert \queryLLM{$t_c$, openapi}\;
    \KwRet \Openapi\;
}
\end{algorithm}
\end{minipage}

To create the queries, we rely on \texttt{createQueries}~(40), with the set of OpenAPIs (\textit{openapis}), the list of endpoints $e_\text{all}$, and the number of queries to be created $n_q$ as parameters.
Starting with an empty list \textit{queries}~(41), we create the $n_q$ queries one by one~(42-47).
Thereby, we select a random subset $e_\text{expected}$ of $e_\text{all}$, where the cardinality of $e_\text{expected}$ is normally distributed~(43).
We set $\mu=5$ and $\sigma=2$, which is about 10\% of all endpoints within the domain.
To create the natural language query (\textit{query}), we invoke \texttt{createQuery} with the OpenAPIs (\textit{openapis}) and the list of expected endpoints ($e_\text{expected}$)~(44).
Once the \textit{query} is created, we check the similarity using OpenAI's \texttt{text-embeddings-3-large} embedding model with the similarity threshold $s_\text{threshold}=0.8$~(45).
If the \textit{query} exceeds the threshold, we discard it and start over from~(43).
Otherwise, we add the \textit{query} to \textit{queries}~(46) and continue with the next one~(42).

\smallskip
\noindent
\begin{minipage}{\linewidth}
\makeatletter
\@twocolumnfalse
\makeatother
\begin{algorithm}[H]
\caption{Create Query}\label{alg:create_queries}
\DontPrintSemicolon
\SetKw{KwTo}{in}
\SetKw{KwAssert}{assert}
\SetKwComment{Comment}{$\triangleright$\ }{}
\SetCommentSty{textrf}
\SetFuncSty{textrf}
\SetNlSty{}{\color{gray}}{}
\SetKwFunction{createQueries}{createQueries}
\SetKwFunction{createQuery}{createQuery}
\SetKwFunction{similarity}{similarity}
\SetKwFunction{queryLLM}{queryLLM}
\SetKwFunction{set}{set}
\SetKwFunction{checkNecessary}{checkNecessary}
\SetDataSty{itshape}
\SetKwData{Queries}{queries}
\SetKwData{Query}{query}
\SetKwData{Endpoints}{endpoints}
\SetKwData{Necessary}{necessary}
\SetKwProg{Fn}{function}{}{end}
\setcounter{AlgoLine}{38}
$s_{\text{threshold}}\gets$ similarity threshold\;
\Fn{\createQueries{openapis, $e_{\text{all}}$, $n_q$}}{
    \Queries $\gets$ []\;
    \For{$i\gets 1, n_q$}{
        $e_{\text{expected}}\gets$ random from $e_{\text{all}}$\;
        \Query $\gets$ \createQuery{openapis, $e_{\text{expected}}$}\;
        \KwAssert \similarity{\Query, \Queries} $<$ $s_{\text{threshold}}$\;
        \Queries.append((\Query,$e_{\text{expected}}$))\;
    }
    \KwRet \Queries\;
}
$t_q\gets$ template create query solution endpoints pair\;
$t_f\gets$ template list further endpoints\;
\Fn{\createQuery{openapis, $e_{\text{expected}}$}}{
    $q\gets$ \queryLLM{$t_q$, openapis, $e_{\text{expected}}$}\;
    $e_{\text{further}}\gets$ \queryLLM{$t_f$, openapis, $q$, $e_{\text{expected}}$}\;
    $e_{\text{extended}}\gets$ \set{$q$.endpoints} $|$ \set{$e_{\text{expected}}$}\;
    $e_{\text{necessary}}\gets$ \checkNecessary{openapis, $q$, $e_{\text{extended}}$}\;
    \KwAssert $e_{\text{necessary}}=e_{\text{expected}}$\;
    \KwRet $q$\;
}
$t_n\gets$ template check endpoint necessary\;
\Fn{\checkNecessary{openapis, $q$, $e_\text{extended}$}}{
    $e_{\text{necessary}}\gets$ []\;
    \For{$endpoint$ \KwTo $e_\text{extended}$}{
        \Necessary $\gets$ \queryLLM{\\\nonl\qquad$t_n$, openapis, $q$, $e_{\text{extended}}$, endpoint}\;
        \lIf{\Necessary}{
            $e_{\text{necessary}}$.append(endpoint)
        }
    }
    \KwRet $e_{\text{necessary}}$\;
}
\end{algorithm}
\end{minipage}

To create the queries, we rely on \texttt{createQueries}~(40), with the set of OpenAPIs (\textit{openapis}), the list of endpoints $e_\text{all}$, and the number of queries to be created $n_q$ as parameters.
Starting with an empty list \textit{queries}~(41), we create the $n_q$ queries one by one~(42-47).
Thereby, we select a random subset $e_\text{expected}$ of $e_\text{all}$, where the cardinality of $e_\text{expected}$ is normally distributed~(43).
We set $\mu=5$ and $\sigma=2$, which is about 10\% of all endpoints within the domain.
To create the natural language query (\textit{query}), we invoke \texttt{createQuery} with the OpenAPIs (\textit{openapis}) and the list of expected endpoints ($e_\text{expected}$)~(44).
Once the \textit{query} is created, we check the similarity using OpenAI's \texttt{text-embeddings-3-large} embedding model with the similarity threshold $s_\text{threshold}=0.8$~(45).
If the \textit{query} exceeds the threshold, we discard it and start over from~(43).
Otherwise, we add the \textit{query} to \textit{queries}~(46) and continue with the next one~(42).

For the creation of a single query, we invoke \texttt{createQuery} with the list of OpenAPIs (\textit{openapis}) and the list of expected endpoints $e_\text{expected}$~(52).
Therefore, we invoke the \gls{llm} (\texttt{queryLLM}) with the template $t_q$ for creating a natural language query, the \textit{openapis}, and $e_\text{expected}$.
The result is the natural language query $q$~(53).
To validate whether the $q$ conforms with $e_\text{expected}$, we again invoke the \gls{llm} (\texttt{queryLLM}) with the template $t_f$, the \textit{openapis}, and $e_\text{expected}$ to list the endpoints $e_\text{further}$, which are necessary to fulfill $q$, but are not in $e_\text{expected}$~(54).
We or the sets $e_\text{expected}$ and $e_\text{further}$ to $e_\text{extended}$~(55).
For each of the endpoints in $e_\text{extended}$, we check if it is genuinely required to fulfill $q$ by calling \texttt{checkNecessary} with the \textit{openapis}, $q$, and the whole list $e_\text{extended}$ to check interdepence.
\texttt{checkNecessary} return the list of genuinely necessary endpoints $e_\text{necessary}$~(56).
If there is a mismatch between $e_\text{necessary}$ and $e_\text{expected}$, we prompt the \gls{llm} in a chat-based manner~(57), i.e., in a question-answer style, with the expected endpoints $e_\text{expected}$, the additional endpoints $e_\text{necessary} \setminus e_\text{expected}$, and the absent endpoints $e_\text{expected} \setminus e_\text{necessary}$ in the response message to the \gls{llm} to improve the prompt and continue with~(54).

The checking for necessity is encapsulated in \texttt{checkNecessary} with the OpenAPIs (\textit{openapis}), the query $q$, and the list of expected to be necessary endpoints $e_\text{extended}$ as parameters~(61).
The idea is to filter out unnecessary endpoints from $e_\text{extended}$.
To realize this, we start with the empty list $e_\text{necessary}$ to store the actually required endpoints to fulfill $q$~(62).
For each of the endpoints in $e_\text{extended}$~(63), we query the \gls{llm} (\texttt{queryLLM}) with the template $t_n$, the \textit{openapis}, $q$, all endpoints $e_\text{extended}$, and the current \textit{endpoint} in a query-answer style to determine whether the current endpoint is required to fulfill $q$~(64).
The \gls{llm}, thereby, returns a simple \enquote{Yes} or \enquote{No} for each endpoint.
If the \gls{llm} returns \enquote{Yes}, we add the endpoint to $e_\text{necessary}$~(65).
This is because stating required endpoints~(54) can reveal many unrelated endpoints, and the query-answer style helps the \gls{llm} to focus as the \gls{llm} might be confused when evaluating all endpoints at once.

\Cref{alg:create_benchmark,alg:create_queries} guarantee that exactly $n_s=5$ services are created, each with exactly $n_e=10$ endpoints, resulting in an OpenAPI complying with the standard.
They create precisely $n_q=10$ queries.
Additionally, they ensure with high probability through the application of an \gls{llm} that the services are non-generic, reasonable services within the expected \gls{gics} domain and that the query can be fulfilled using the given set of services using the stated set of endpoints exclusively and through using an embedding model that the queries within one domain do not exceed a similarity threshold.

\begin{figure*}%
    \subfloat[Top 10 Candidates by Recall and Precision. NV is the Nvidia model, and OAI is the OpenAI model. ES represents Endpoint Split with token chunking with the overlap $l$ in parentheses.]{%
        \begin{adjustbox}{width=\columnwidth,valign=t}%
            \begin{tabular}{r|l|l|l|l|l|l}%
                & \multicolumn{3}{c|}{Recall} & \multicolumn{3}{c}{Precision} \\
                \makecell[c]{\#} & Model & Strategy & K & Model & Strategy & K \\
                \hline
                1. & NV & Summary & 20 & NV & Summary & 5 \\
                2. & NV & ES JSON & 20 & NV & ES JSON & 5 \\
                3. & NV & ES (20) & 20 & NV & ES (0) & 5 \\
                4. & NV & ES (0) & 20 & NV & ES (20) & 5 \\
                5. & NV & ES Thin & 20 & NV & ES Thin & 5 \\
                6. & NV & CRAFT & 20 & NV & ES Field & 5 \\
                7. & OAI & ES JSON & 20 & OAI & ES JSON & 5 \\
                8. & NV & ES Field & 20 & OAI & ES (20) & 5 \\
                9. & OAI & ES Thin & 20 & OAI & ES (0) & 5 \\
                10. & OAI & ES (20) & 20 & OAI & ES Thin & 5%
            \end{tabular}%
            \label{fig:crossdomain:metrics}%
        \end{adjustbox}%
    }%
    \hfill%
    \subfloat[Pareto Front Analysis of Recall and Precision as Scatterplot. Model Color-Coded. $k$ Shape-Coded.]{%
        \includegraphics[width=\columnwidth,valign=t]{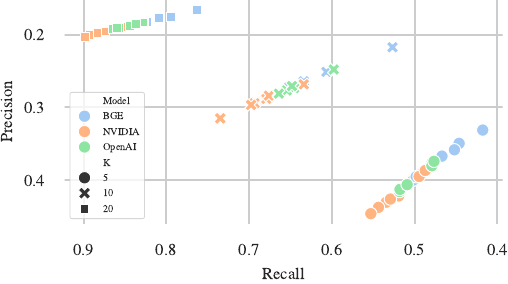}%
        \label{fig:crossdomain:pareto}%
    }%
    \caption{Cross-Domain Average Analysis}%
    \label{fig:crossdomain}%
\end{figure*}

\subsection{Evaluation Methodology}

For the result evaluation, we consider the service discovery with \gls{rag} as a hyperparameter tuning problem with $k$ for the top-$k$ selection of candidates, the model $m$, and the chunking strategy $s$ as independent variables.
Further, we consider the domains as independent datasets.
The methodology follows:
\\[1.2ex]
\textbf{(1) Performance criteria:} We define the dependent variables recall and precision as the performance criteria metrics because we are interested in how many correct endpoints we retrieve.
        We are not interested in the ranking of the endpoints because we assume that the incorrect endpoints are filtered out in a later stage, i.e., we do not consider other metrics like Mean Reciprocal Rank, which weigh positioning and Hit@k, which does not consider the number of correct results.
\\[1.2ex]
\textbf{(2) Candidate set:}
        As embedding models $M$, we employ OpenAI's \texttt{text-embedding-3-large}~\cite{openai2024embeddings} as one of the currently leading proprietary models.
        As open-source models, we utilize BAAI/\theosmodelsmall~\cite{xiao2023bge}, which is relatively small while still producing reasonable results, allowing the model to be executed on commonly available hardware like laptops, and Nvidia's \theosmodellarge~\cite{lee2024nv} as one of the leading open-source models.
        For the parameter $k$, we set $K=\{5,10,20\}$ as these are the multiples of $\mu=5$ of our normal distribution, and for the chunking strategy $s$, we use the chunking strategies $S$ as defined in \Cref{tab:approaches}.
        This results in the candidate set $C=\{(m,k,s)|m \in M,k \in K,s \in S\}$.
\\[1.2ex]
\textbf{(3) Domain-dependent datasets:} For each $c \in C$ and each domain, we execute \thebenchmark, resulting in a set of independent result sets.
\\[1.2ex]
\textbf{(4) Cross-domain average:} As we are interested in the candidate that performs best across all domains, we compute the performance criteria metrics as an average across all domains for each candidate $c \in C$ using the domain-dependent datasets. We weigh each domain equally as we consider each domain as equally important.
\\[1.2ex]
\textbf{(5) Stability:} We analyze the standard deviation across domains to determine whether there is a candidate $c \in C$ that performs slightly worse but performs more stable.
          We compute the standard deviation of the recall across all domains and $k$ and separate it by model $m$ and chunking strategy $s$.
\\[1.2ex]
\textbf{(6) Significance:} We perform the Friedman test as we do not assume normality to resolve if candidate differences are statistically significant across domains.
          We evaluate the test for each domain and across all domains by model $m$ and $k$.

\subsection{Experimental Results on \thebenchmark}

We first focus on the chunking strategies with endpoint split as splitting strategy, i.e., endpoint split splitting with token chunking, remove examples, relevant fields, JSON split token chunking, query, summary, and CRAFT, as these always reveal exactly one endpoint per chunk.
Then, we cross-validate it with the remaining document and JSON split approaches.

\begin{figure*}
    \includegraphics[width=\linewidth]{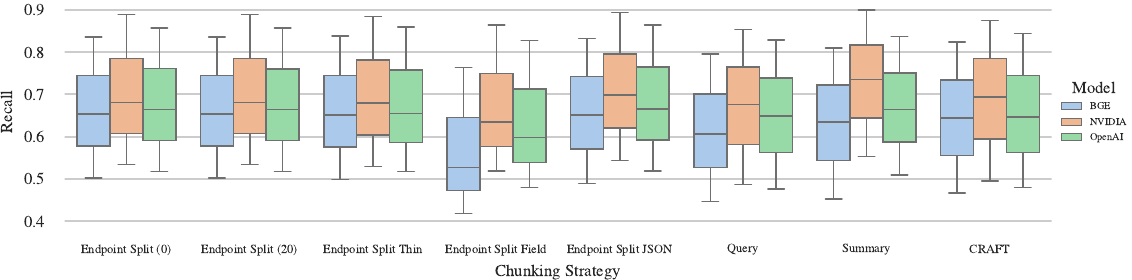}
    \caption{Recall by Chunking Strategy as Boxplots Grouped by Model for $k=20$. Model Color-Coded.}
    \label{fig:model_comparison}
\end{figure*}

\begin{figure*}%
    \subfloat[Recall by Chunking Strategy for $k=20$ Grouped by Model with Standard Deviation as Error Bars.]{%
        \includegraphics[width=\columnwidth,valign=b]{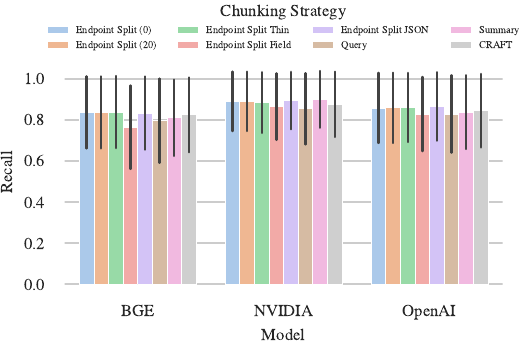}%
        \label{fig:stability:chunking_strategy}%
    }%
    \hfill%
    \subfloat[Standard Deviation by Mean Recall for All Candidates. Color-Coded by $k$. Coefficient of Variation (CV) Size-Coded.]{%
        \includegraphics[width=\columnwidth,valign=b]{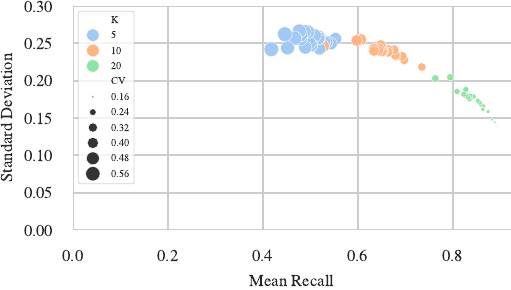}%
        \label{fig:stability:cv}%
    }%
    \caption{Statistical Stability Analysis of the Candidates.}%
    \label{fig:stability}%
\end{figure*}

The cross-domain average is shown in \Cref{fig:crossdomain}.
The figure is split into two subfigures.
\Cref{fig:crossdomain:metrics} presents the top 10 candidates by recall and precision.
The top 10 candidates in recall all have $k=20$, and for precision, $k=5$.
Also, the top six candidates use the Nvidia model.
The remaining candidates either use the Nvidia or the OpenAI model, revealing the superiority of these models over the BGE model.
Both metrics list the summary approach with the Nvidia model as the leading chunking strategy but with different $k$.

To further refine the results, we perform a Pareto front analysis for recall and precision shown in \Cref{fig:crossdomain:pareto} as a scatterplot.
The abscissa shows the recall and the ordinate the precision with an inverse scale, i.e., the best result is in the origin, and the closer to the origin, the better the result.
We can determine three distinct clusters defined by the $k$-value, unveiling that with a higher $k$, the recall increases while the precision drops and vice versa.
In total, we see the three Nvidia summary candidates as Pareto optimal results, one for each cluster.
Nevertheless, differences between the chunking strategies are not apparent within a cluster.
Further refinement through color-coding of the model reveals that the Nvidia model outperforms the OpenAI model, which in turn outperforms the BGE model.
Still, the differences are minor compared to the $k$ clusters.

To further analyze and compare the different models, we create a boxplot chart of the recall for $k=20$ in \Cref{fig:model_comparison}.
We group the boxplots by chunking strategy and color-coding the model.
For each chunking strategy, the median recall of the Nvidia model is above the OpenAI model, which, in turn, always performs better than the BGE model.
The Nvidia model performs exceptionally well with the summary approach, while the BGE model performs poorly with the endpoint split field approach.
Otherwise, there are no differences visible between the chunking strategies.
Still, the interquartile range overlaps, requiring further analysis to determine an obvious winner.
These results align with the results of the MTEB leaderboard\footnote{\url{https://huggingface.co/spaces/mteb/leaderboard}}, which ranks current embedding models.

\inlineparagraph{Stability Analysis}
Another factor is stability, which we analyze in \Cref{fig:stability}.
The first analysis segment is \Cref{fig:stability:chunking_strategy}, which shows average recall by chunking strategy as a bar chart grouped by the model for $k=20$.
Stability is shown as error bars of the standard deviation, which is between 13-25\%.
We can determine that the mean is again lower and varies more for the BGE model than the other two.
Also, the figure shows that all models, while revealing significant performance differences, expose a good overall performance.
Yet, for all three models, the differences in mean between the chunking strategies seem small.
Considering the error bar, the mean differences seem minor as the variance outweighs these.
The standard deviation over mean recall for all candidates in a scatterplot, with the coefficient of variance $\text{CV} = \frac{\text{standard deviation}}{\text{mean}}$ as the point size is shown in \Cref{fig:stability:cv}.
$k$ is shown as color, again revealing the clustering in recall comparable to the one seen in \Cref{fig:crossdomain:pareto}.
We can see a tendency for the standard deviation to decrease with a higher $k$ for the clusters.
For $k=5$, the mean standard deviation is about 4\%; for $k=10$, it is about 3.5\%; and for $k=20$, it is around 3\%.
This is also seen in the coefficient of variance.
Within a cluster, there is no clear winning chunking strategy and model.

\begin{figure*}%
    \subfloat[Friedman Test by Domain And Across All Domains for the BGE, the Nvidia (NV), and the OpenAI (OAI) Model and $k=\{5,10,20\}$. Values above the significance level, i.e., $p \ge 0.05$, are marked in bold.]{%
        \begin{adjustbox}{width=0.675\linewidth,valign=t}%
            \begin{tabular}{r|l|l|l|l|l|l|l|l|l}
                & \multicolumn{9}{c}{k} \\
                & \multicolumn{3}{c|}{5} & \multicolumn{3}{c|}{10} & \multicolumn{3}{c}{20} \\
                Domain & BGE & NV & OAI & BGE & NV & OAI & BGE & NV & OAI \\
                \hline
                Communication Services & 0.00 & 0.02 & 0.01 & 0.00 & 0.00 & 0.00 & 0.00 & 0.00 & 0.00 \\
                Consumer Discretionary & \textbf{0.10} & 0.01 & 0.03 & 0.02 & 0.00 & \textbf{0.06} & 0.00 & \textbf{0.06} & \textbf{0.98} \\
                Consumer Staples & 0.00 & \textbf{0.43} & \textbf{0.10} & 0.00 & 0.01 & 0.00 & \textbf{0.40} & 0.04 & \textbf{0.61} \\
                Energy & 0.00 & 0.00 & 0.00 & 0.00 & 0.00 & 0.01 & 0.00 & \textbf{0.06} & 0.01 \\
                Financials & 0.00 & 0.00 & 0.03 & 0.02 & \textbf{0.12} & 0.01 & \textbf{0.06} & 0.00 & \textbf{0.73} \\
                Health Care & \textbf{0.06} & 0.00 & \textbf{0.27} & 0.00 & 0.00 & 0.01 & 0.00 & 0.00 & 0.03 \\
                Industrials & 0.00 & \textbf{0.35} & \textbf{0.24} & 0.00 & 0.00 & 0.00 & 0.00 & \textbf{0.30} & 0.01 \\
                Information Technology & 0.01 & \textbf{0.54} & \textbf{0.27} & 0.00 & 0.00 & 0.00 & 0.00 & \textbf{0.31} & 0.00 \\
                Materials & 0.00 & \textbf{0.50} & 0.01 & 0.00 & 0.00 & 0.03 & 0.05 & 0.00 & \textbf{0.21} \\
                Real Estate & 0.00 & 0.00 & 0.00 & 0.00 & 0.00 & 0.00 & 0.00 & 0.00 & 0.00 \\
                Utilities & 0.00 & 0.00 & 0.02 & 0.00 & 0.00 & 0.04 & 0.03 & 0.00 & \textbf{0.13} \\
                \hline
                All & 0.00 & 0.00 & 0.00 & 0.00 & 0.00 & 0.00 & 0.00 & 0.00 & 0.00
            \end{tabular}%
            \label{tab:friedman}%
        \end{adjustbox}%
    }%
    \hfill%
    \subfloat[Ascending Token Count per Chunk by Chunking Strategy Averaged over $k={5,10,20}$]{%
    \begin{adjustbox}{width=0.3\linewidth,valign=t}%
            \begin{tabular}{r|r}
                Chunking Strategy & \#Token \\
                \hline
                Whole Document (100, 0) & 34.93 \\
                Whole Document (100, 20) & 35.16 \\
                Whole Document (200, 20) & 56.87 \\
                Whole Document (200, 0) & 56.96 \\
                JSON (100, 0) & 97.46 \\
                JSON (100, 20) & 97.76 \\
                \hline
                Endpoint Split Thin & 121.94 \\
                Endpoint Split JSON & 131.85 \\
                Endpoint Split Field & 152.33 \\
                Endpoint Split (0) & 155.31 \\
                Endpoint Split (20) & 155.32 \\
                Query & 155.70 \\
                CRAFT & 157.41 \\
                Summary & 161.29 \\
            \end{tabular}%
            \label{tab:token_count}%
        \end{adjustbox}%
    }%
    \caption{Friedman Test for the Endpoint Split and Token Count per Chunk for All Chunking Strategies from \Cref{tab:approaches}.}%
    \label{fig:friedman_token}%
\end{figure*}

\inlineparagraph{Significance Analysis}
To compare the chunking strategies, if there is a significant difference between the individual chunking strategies, we employ the Friedman test as it can compare multiple groups without requiring a normal distribution in the measurements.
We perform the Friedman test for each domain and over all domains and set the significance level to 5\%.
The results are shown in \Cref{tab:friedman} split by $k$ and model $m$.
Entries exceeding the significance level are marked in bold.
We can extract that there are some differences within chunking strategies within specific domains, but these average out over all domains.
This can be through the training data of the embedding model or through the increasing variance when condensing a more extensive set of measurements.
Therefore, we can assume that there is no significant difference between chunking strategies over all domains, but individual cases may be considered when choosing the chunking strategy.

\begin{figure*}%
    \subfloat[Scatterplot of Recall and Precision of all Candidates with the Whole Document and JSON Splitting Strategy. The Parameters Chunk Size $s$ and Overlap $l$ are in Parantheses $(s,l)$. Chunking Strategy Color-Coded. $k$ Shape-Coded.]{%
        \includegraphics[width=\columnwidth,valign=b]{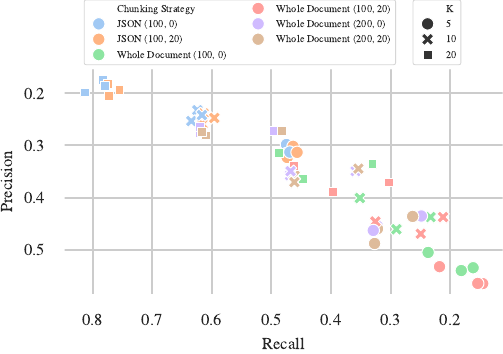}%
        \label{fig:non_es:non_es}%
    }%
    \hfill%
    \subfloat[Comparison to Endpoint Split-Based Chunking Strategies as a Scatterplot Formated like \Cref{fig:non_es:non_es}. Candidates are Limited to the Nvidia Model. The Summary Chunking Strategy is Added for Comparison.]{%
        \includegraphics[width=\columnwidth,valign=b]{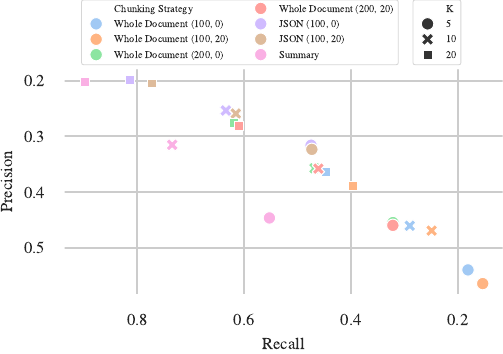}%
        \label{fig:non_es:non_es_summary}%
    }%
    \caption{Evaluation of Chunking Strategies with Non-Endpoint Split Splitting, i.e., Whole Document and JSON.}%
    \label{fig:non_es}%
\end{figure*}

\inlineparagraph{Analysis of Multi-Endpoint Chunking Strategies}
Compared to the endpoint split-based chunking strategies, the whole document and JSON approaches do not guarantee that each chunk corresponds precisely to one endpoint, i.e., one chunk can contain multiple endpoints or fragments of these.
In addition, they allow the specification of the chunk size $s$ and the overlap $l$ in tokens as they rely on the token chunking refinement.
As the endpoints in the \thebenchmark OpenAPIs are shorter than comparable real-world OpenAPIs through the \gls{llm} generation, we choose small values for $s$ and $l$ to account for more realistic results.
We set $l=\{0,20\}$ and $s=\{100,200\}$ for the whole document and $l=\{0,20\}$ and $s=\{100\}$ for the JSON splitting strategies.
Otherwise, a single chunk can contain numerous endpoints, making evaluation difficult due to incomparability.

First, we examine the token count shown in \Cref{tab:token_count} to overview how much data is considered on average per chunk and chunking strategy.
The chunking strategies and average token count per chunk are shown as sorted ascending by average token count.
For the token counting, we rely on the \enquote{tiktoken} library and select the \enquote{gpt-4o} as the target model.\footnote{\url{https://github.com/openai/tiktoken}}
The average is computed over all models and $k=\{5,10,20\}$.
The whole document splitting approaches produce chunks much smaller than $s$, which can be explained due to the different token counting of LlamaIndex and tiktoken, where tiktoken can combine multiple tokens into one, e.g., considering white spaces and structural elements in the OpenAPI JSON input.
Also, the JSON splitting creates much more dense chunks, i.e., with more tokens per chunk than the whole document approaches.
Further, for our chosen parameters, the endpoint split chunking strategies produce larger chunks on average compared to the whole document and JSON approaches.
A larger $s$ like in real-world cases can reverse this effect as lengthy chunks can be condensed more effectively, e.g., using the summary approach independently of the input length.
Overall, the token count for all strategies is relatively low, allowing us to insert hundreds of chunks into a single prompt.

Considering accuracy, \Cref{fig:non_es} presents the results of the non-endpoint split-based chunking strategies, i.e., whole document and JSON, as scatterplots of recall as inversed abscissa and precision as inversed ordinate.
The left side, \Cref{fig:non_es:non_es}, depicts all candidates of all models with the chunking strategy color-coded and $k$ shape-coded.
Compared to the endpoint split-based chunking strategies shown in \Cref{fig:crossdomain:pareto}, there are no clear clusters or obvious Pareto-optimal strategy.
On the upper left, we can see that the JSON split chunking strategies tend to perform better concerning recall but also reveal low precision.
This can be explained by the implementation of the JSON split algorithm, which densely packs the information from the OpenAPIs, resulting in many endpoints represented in a single chunk, increasing recall while decreasing precision.
Also, the tendency that a higher $k$ leads to a higher recall and a lower precision can be determined by most of the chunking strategies with $k=5$ being in the lower right, with $k=10$ in the center, and $k=20$ in the upper left.
Regarding the whole document splitting, the $s=200$ candidates seem to surpass the $s=100$ candidates in recall with a decrease in precision, which can stem from the inclusion of multiple endpoints in one chunk or the embedding model receiving more information, which results in better similarity.
No prominent difference results from including the overlap $l=20$ compared to $l=0$.

\Cref{fig:non_es:non_es_summary} restricts the candidates to the Nvidia model to ease comparison and adds the summary approach.
What becomes apparent is that the summary approach is Pareto-optimal with a large margin for all $k$.
Nevertheless, the whole document approaches with $s=100$ and $k=5$ reveal a higher precision but with a significant drop in recall.
Due to this high margin, we infer that the endpoint split approaches outperform the whole document and JSON splitting-based approaches in our experiments.
Further research is needed to determine the influence of $s$, especially regarding extensive OpenAPIs.

\inlineparagraph{Summary of Findings}
In summary, the biggest influence on accuracy is $k$.
Therefore, we recommend practitioners choose the highest $k$ possible for their use case to achieve the highest recall.
The second most considerable influence is on the embedding model $m$.
As the differences are not as prominent, we recommend the Nvidia model if the highest accuracy is needed, the OpenAI model as the model if the practitioners are already familiar with the OpenAI tooling, and the BGE model if a small resource footprint is required.
Regarding the chunking strategy, endpoint split-based approaches outperform whole document or JSON split-based approaches.
Within endpoint split-based approaches, there is no significant difference across all domains.
Therefore, we recommend choosing the simplest one to implement.
Suppose the \gls{rag} should be employed in a very special domain; a specific chunking strategy might be beneficial, which should be determined individually for the actual case.

\subsection{Experimental Results on RestBench}

To evaluate \therag in a real-world setting, we employ the RestBench benchmark in addition to \thebenchmark, covering the Spotify and TMDB OpenAPI specifications~\cite{song2023restgpt}.
The services of Restbench, with 40 endpoints for Spotify and 54 for TMDB, are much more complex than usual \gls{soc} case studies containing usually just three to seven endpoints~\cite{pesl2024uncovering}.
Nevertheless, it only covers the communication services domain, with 57 queries for Spotify and 100 for TMDB.
Thus, it is significantly smaller than \thebenchmark.
Also, the queries can be much more vague.
For \thebenchmark, we ensure that the query precisely aligns with the expected endpoints through an iterative process, resulting in an unambiguous query.
This is not the case for RestBench.
For evaluation, we perform the same steps described in \Cref{sec:socbenchd} for \thebenchmark.

\begin{figure}
    \includegraphics[width=\columnwidth]{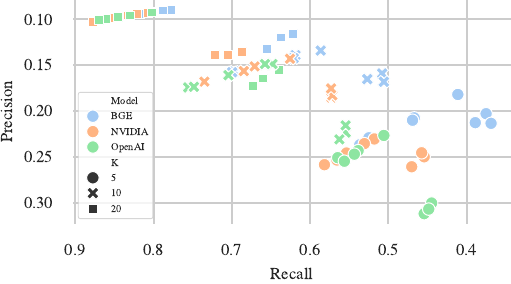}
    \caption{RestBench Pareto Front Analysis of Recall and Precision as Scatterplot. Model Color-Coded. $k$ Shape-Coded.}
    \label{fig:restbench_pareto}
\end{figure}

Like \Cref{fig:crossdomain:pareto}, \Cref{fig:restbench_pareto} shows the Pareto front analysis of the endpoint split-based chunking strategies.
The $k$ clusters are less evident due to a higher variance.
Still, $k=20$ tends to express higher recall and lower precision than $k=10$, resulting in a higher recall but lower precision than $k=5$.
The Nvidia and OpenAI models distinctly outperform the BGE model.
There is no clear dominance between the Nvidia and the OpenAI model.
Performing the Friedman test again reveals no significant difference between the endpoint-split-based chunking strategies.

\begin{figure*}%
    \subfloat[Scatterplot of Recall and Precision of all Candidates with the Whole Document and JSON Splitting Strategy. The Parameters Chunk Size $s$ and Overlap $l$ are in Parantheses $(s,l)$. Chunking Strategy Color-Coded. $k$ Shape-Coded.]{%
        \includegraphics[width=\columnwidth,valign=t]{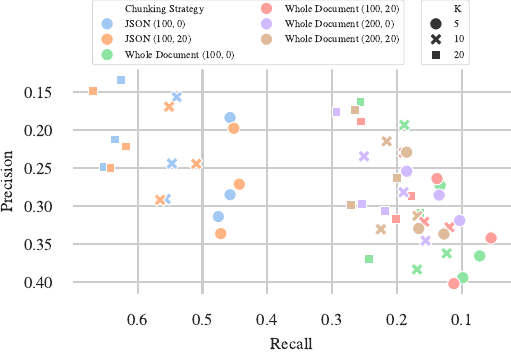}%
        \label{fig:restbench_non_es:non_es}%
    }%
    \hfill%
    \subfloat[Comparison to Endpoint Split-Based Chunking Strategies as a Scatterplot Formated like \Cref{fig:restbench_non_es:non_es}. Candidates are Limited to the Nvidia Model. The Summary Chunking Strategy is Added for Comparison.]{%
        \includegraphics[width=\columnwidth,valign=t]{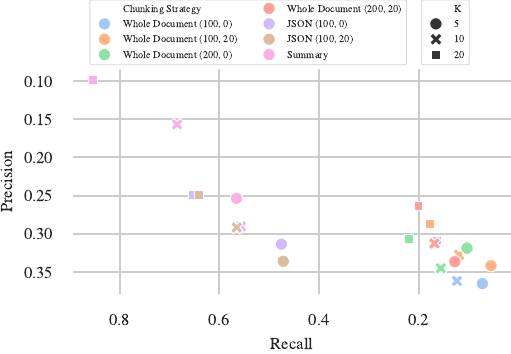}%
        \label{fig:restbench_non_es:summary}%
    }%
    \caption{RestBench Evaluation of Chunking Strategies with Non-Endpoint Split Splitting, i.e., Whole Document and JSON. Formatting as in \Cref{fig:non_es}.}%
    \label{fig:restbench_non_es}%
\end{figure*}

\begin{figure*}%
    \subfloat[\thebenchmark]{%
        \includegraphics[width=\columnwidth,valign=t]{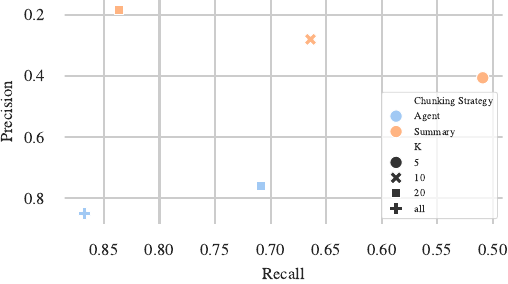}%
        \label{fig:pareto_agent:socbenchd}%
    }%
    \hfill%
    \subfloat[RestBench]{%
        \includegraphics[width=\columnwidth,valign=t]{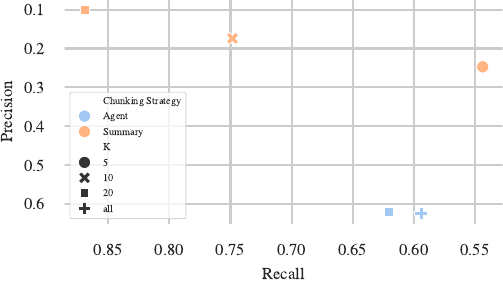}%
        \label{fig:pareto_agent:restbench}%
    }%
    \caption{Pareto Front Analysis of the \theagent as a Scatterplot. Agent Results are in Blue. The Summary Chunking Strategy is in Orange. $k$ is Shape-Coded.}%
    \label{fig:pareto_agent}%
\end{figure*}

As for the whole document and JSON splitting strategy-based chunking strategies, we perform the same experiments in \Cref{fig:restbench_non_es} for RestBench as in \Cref{fig:non_es} for \thebenchmark.
In \Cref{fig:restbench_non_es:non_es}, we can see a sharp distinction in recall between the whole document and the JSON approaches, with the JSON approaches performing significantly better.
Among the JSON candidates are three diagonal clusters representing the different models, with the Nvidia model performing best and the BGE model performing worst.
Within the cluster, there is again the relationship with an increase in $k$, resulting in an increase in recall and a drop in precision.
This correlation is also visible for the whole document approach but with a minor difference in the recall.
When plotted with models visible, for the whole document approaches, the OpenAI model performs best in precision, followed by the Nvidia model, then the BGE model.
This can result from special training data.
There is no apparent relation between the chunking strategies, the model, and recall.

By adding the summary approach to compare the endpoint split-based approaches with the whole document and JSON approaches, we can determine that the summary approach outperforms the other approaches with the same $k$ in recall.
Yet, the whole document and JSON approach surpass the summary approach in precision, which can result from multiple chunks being retrieved for the same endpoint.
This increases precision as the cardinality of the set of retrieved endpoints decreases.

Overall, the RestBench results reinforce the \thebenchmark results while revealing a higher variance.
Further research is needed on chunk size $s$ and overlap $l$'s influence on the whole document and the JSON splitting-based strategies.

\subsection{\theagent}

To evaluate whether the results can be further improved by employing the \theagent as a means to utilize the \gls{llm} agent reasoning capabilities, we execute \thebenchmark and RestBench and compare the \theagent results with the standalone \gls{rag}.
\Cref{fig:pareto_agent:socbenchd} shows the \thebenchmark results for \theagent and, for comparison, the summary approach for the OpenAI model as a scatterplot over recall as the inversed abscissa and precision as the inversed ordinate.
For \theagent, we employ $k=\{20,\text{all}\}$ as the maximum number of elements, which are retrieved per query to the \gls{rag} tool that the agent can use.
The agent can then refine the query and filter the returned endpoints.
$k=20$ is used for comparison with the standalone counterparts.
$k=\text{all}$ reveals always all endpoints.
The results show that the agent with $k=\text{all}$ outperforms the standalone summary \gls{rag} approach in recall and precision.
For both $k=\{20,\text{all}\}$, the agent increase precision.
When comparing $k=20$, the agent increases precision but seems to filter out relevant endpoints, resulting in a recall drop.
Comparing the $k=\text{all}$ agent with the $k=20$ summary reveals a minor increase in recall (87\% vs. 84\%).
Nevertheless, the agent cannot identify all relevant endpoints, even when fed with all endpoints.

We present the RestBench results of the \theagent in \Cref{fig:pareto_agent:restbench}.
The setup equals the \thebenchmark results from \Cref{fig:pareto_agent:socbenchd}.
Compared to the \thebenchmark results, the summary results are similar in recall with a lower precision.
For $k=\{20,\text{all}\}$, the agent again improves precision significantly.
For $k=20$, like in \Cref{fig:pareto_agent:restbench}, the agent lowers the recall, i.e., it filters out endpoints too restrictively.
The main difference is that for $k=\text{all}$, the agent performs notably worse in recall and reveals a similar precision compared to $k=20$.
This can be due to more endpoints and more similar endpoints added to the prompt, making it harder for the model to determine relevant ones.

In conclusion, the benefit of the \theagent is the increased precision, which reveals most of the necessary information while not exposing unrelated information, i.e., the token count is minimal.
The drawback is the additional effort in calling another \gls{llm} and the decrease in recall due to too strict filtering of the \gls{llm}.
Further research is needed to improve the \theagent, e.g., with additional reasoning components to mitigate the filtering issues and confusion due to many endpoints.
In practice, a pragmatic approach might be employing the summary approach with an increase of $k$ to achieve comparable results in the recall, saving the additional effort for the agent.

\subsection{Discussion}

To answer RQ1, we implemented the novel \thebenchmark benchmark based on \gls{gics}, which ensures covering the most relevant domains.
It shows that \therag and the \theagent can retrieve large portions of relevant data while not revealing all relevant information in all cases.

In relation to RQ2, we showed the effectiveness using \thebenchmark and the RestBench benchmark.
Overall, the prototype exhibited the ability to adequately reduce the token size to fit into the \gls{llm} context size while maintaining most of the relevant information.
Regarding the chunking strategies, endpoint split-based chunking strategies achieve favorable accuracies.
There is no significant difference within the endpoint-split chunking strategies.
Limitations for postprocessing are primarily that the \gls{rag} results may not contain all relevant information.

For RQ3, using the summary approach, the \theagent showed improved precision.
Further research is needed to improve the decline in recall due to too strict filtering of the \gls{llm}.
While covering various domains, \thebenchmark is limited to the generated services.
These may be less extensive than comparable real-world services through the \gls{llm} generation process, making token count evaluation less robust.
Further, our real-world evaluation relies on RestBench, which consists of only two services within one domain.
This calls for additional real-world data on endpoint discovery.

Additionally, we rely on pre-trained general-purpose embedding models.
These are trained to perform all kinds of similarity matching and may highlight in our use case insignificant pieces of information.
To further improve performance, fine-tuning, a custom embedding model, or a similarity threshold can be applied to match endpoints to tasks more precisely.

Further, our experiments assume that schemas are inlined in the endpoint specification.
In practice, schemas may be outsourced and linked by a reference.
Further research is needed to determine the best way to process these.

\therag systems in practice may operate on much larger datasets than the ones from \thebenchmark or RestBench.
For the data processing, we rely on standard \gls{rag} implementations like LlamaIndex, which are already designed to operate on large amounts of data.

The applicability of the \therag depends on the availability of service documentation.
We try to mitigate this issue by relying on widely adopted OpenAPI specifications, but this might not be valid for all domains.
A solution to consider is automatically generating service documentation using an \gls{llm}.
Another factor influencing the discovery is the quality of the OpenAPI specifications.
The discovery may fail without descriptions, meaningful naming, or erroneous information.
This is not an issue of the approach, as a human developer would face the same problem, but it highlights the importance of high-quality documentation.

Besides capabilities of the \gls{rag} system, resource consumption is a major issue in \gls{llm}-based systems.
The \therag only uses embedding models.
These are much more efficient than \glspl{llm}, resulting in costs in fractions of a cent per query.
In contrast, the \theagent requires significantly more resources due to relying on an \gls{llm} for the endpoint selection.

\section{Concluding Remarks}
\label{sec:conclusion}

The service discovery challenge is central to  \gls{soc}.
With the application of automated \gls{llm}-based service composition approaches, the \gls{llm} input context limitations have become prominent, as the entire service documentation often does not fit into the input context, requiring the preselection of relevant information.
To address this issue, we proposed an \therag, which facilitates search based on state-of-the-practice OpenAPIs and reduces the input token size.
Further, we show an advanced integration through a \theagent, which can retrieve service details on demand to reduce the input token count further.
Our evaluation based on our novel general-purpose benchmark \thebenchmark and the RestBench benchmark shows that our approach is viable and efficient.


\begin{IEEEbiography}[{\includegraphics[width=1in,height=1.25in,clip,keepaspectratio]{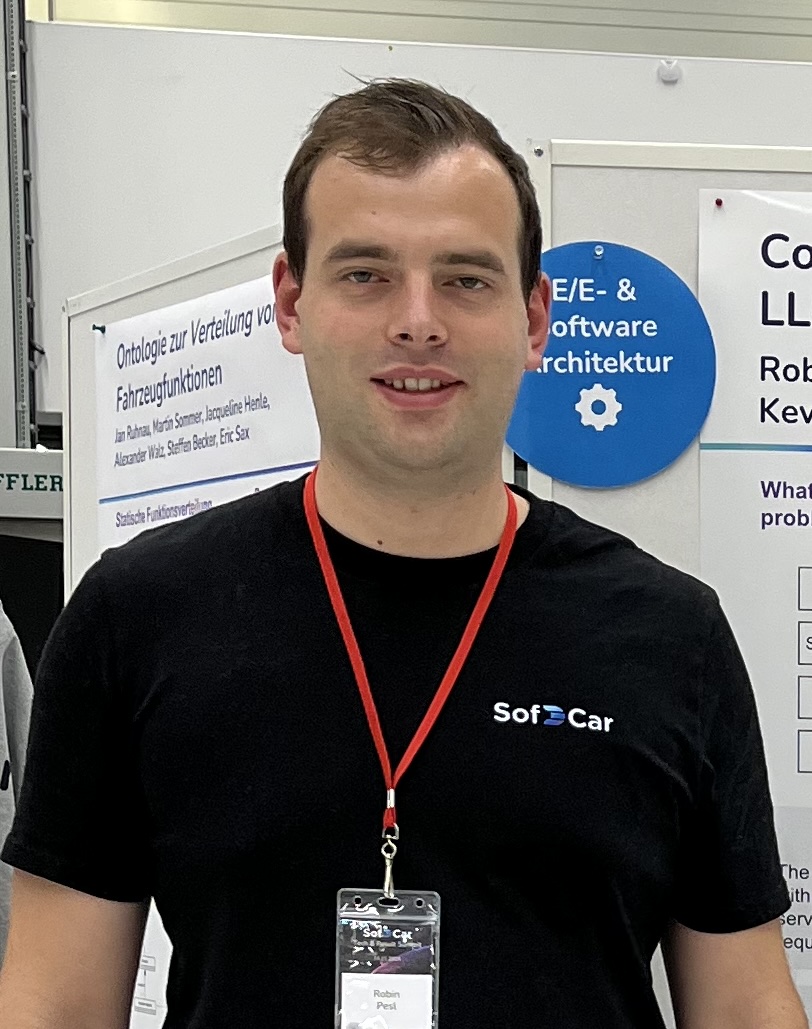}}]{Robin D. Pesl} is a Ph.D. student at the Institute of Architecture of Application Systems at the University of Stuttgart with the focus on the application of \glspl{llm} in Service Computing.
He accomplished his Bachelor's degree in Computer Science at the Cooperative State University Baden-Württemberg as a dual study in cooperation with SAP and his Master's degree at the University of Stuttgart while working part-time at SAP in the SAP HANA Spatial team as a software engineer.
\end{IEEEbiography}

\begin{IEEEbiography}[{\includegraphics[width=1in,height=1.25in,clip,keepaspectratio]{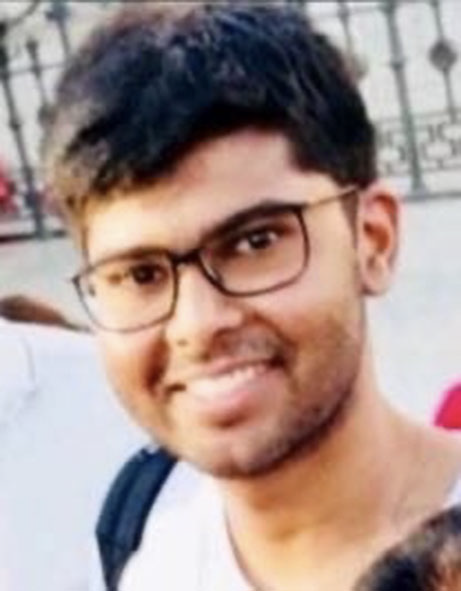}}]{Jerin G. Methew}
received his Ph.D. as a National Doctorate in AI at Sapienza University of Rome.
He pursued his Bachelor's degree in Computer Engineering at Roma Tre University in 2017 and then earned a Master's in Computer Engineering in 2020 at the same university.
His research interests are in data management-related topics, including entity resolution, knowledge graphs, assessment of the fairness of rankings, and applying NLP techniques in these fields.
\end{IEEEbiography}

\begin{IEEEbiography}[{\includegraphics[width=1in,height=1.25in,clip,keepaspectratio]{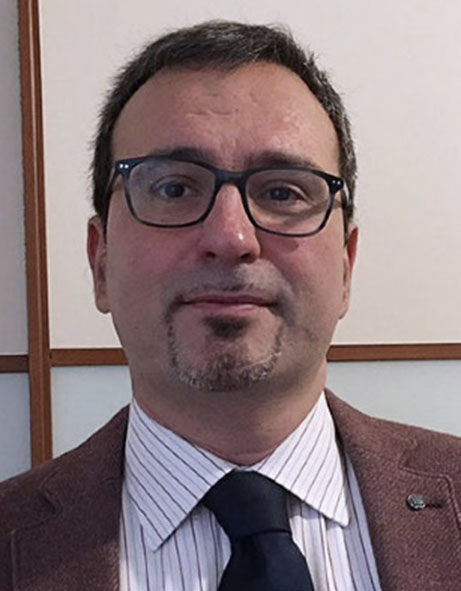}}]{Massimo Mecella}
received the Ph.D. degree in engineering in computer science from the University of Rome \enquote{La Sapienza.}
He is currently a Full Professor at the University of Rome \enquote{La Sapienza,} where he is conducting research, among others, in information systems engineering, software architectures, distributed middleware, and service-oriented computing, focusing on smart applications.
He has various experiences organizing scientific events, e.g., as the General Chair of CAiSE 2019, BPM 2021, and ICSOC 2023.
\end{IEEEbiography}

\begin{IEEEbiography}[{\includegraphics[width=1in,height=1.25in,clip,keepaspectratio]{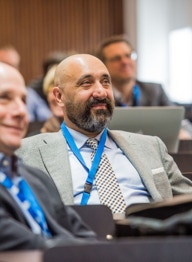}}]{Marco Aiello}
received the Ph.D. degree from the University of Amsterdam, The Netherlands.
He is currently a full Professor of Computer Science and Head of the Service Computing Department at the University of Stuttgart, Germany.
He is an elected member of the European Academy of Sciences and Arts.
His main areas of expertise are in the coordination of cyber-physical systems in complex, dynamic, and uncertain environments.
\end{IEEEbiography}

\vfill

\end{document}